\documentclass[prx,aps,amssymb,twocolumn,noshowpacs,hyperref]{revtex4-1}
\usepackage{color}
\usepackage{graphicx}
\usepackage{hyperref}
\usepackage{amsmath}
\usepackage{latexsym}
\usepackage{amssymb}
\usepackage{mathrsfs}
\usepackage{mathtools}
\usepackage{layout}
\usepackage{verbatim}
\usepackage{multirow}
\usepackage{bm}
\usepackage{amsfonts,epsfig}
\usepackage{textcomp}
\usepackage{diagbox}

\begin{document}

\title{Asymmetric quantum Rabi model, trap-dipole resonance, and quantum gates with optically trapped ultracold polar molecules}

\author{Yan Lu}
\affiliation{Center for Theoretical Physics and School of Physics and Optoelectronic Engineering, Hainan University, Haikou 570228, China}
\author{Xiao-Feng Shi}
\affiliation{Center for Theoretical Physics and School of Physics and Optoelectronic Engineering, Hainan University, Haikou 570228, China}
\begin{abstract}
Optically trapped ultracold polar molecules can have multiple long-lived states for coding quantum information, and can exhibit electric dipole-dipole interactions~(DDI) which enables entanglement generation. The general understanding on the quantized motion~(QM) of molecules in the traps is that it causes fluctuation of DDI. Here, we find that the molecular QM can realize an asymmetric quantum Rabi model, which is of specific importance in the study of fundamental physics. The molecular QM can also lead to an exotic trap-dipole resonance, resulting in excess population loss to uncoupled motional states, and, hence, should be avoided in a general quantum control over polar molecules. To examine the impact of QM on quantum computing based on polar molecules, we introduce two gate protocols, a fast iSWAP gate which can be realized by a global microwave pulse of pulse area smaller than $2\pi$, and a controlled-phase gate with an arbitrary controlled phase, and find that both gates can attain a high fidelity.
\end{abstract}
\maketitle

\section{Introduction}
Ultracold polar molecules in or near their rovibrational ground manifold can provide multiple controllable and addressable ground substates for coding quantum information with second-scale coherence times~\cite{Hepworth_2025}, and, importantly, possess intrinsic molecular-frame electric dipole moments which enable deterministic entanglement generation~\cite{Cornish_2024}. Polar molecules are usually optically trapped during the quantum control by external fields, where the in-trap quantized motion~(QM) of the molecules leads to finite extension of their spatial positions in the traps. Such a QM leads to an uncertainty of the actual distance and orientation between nearby molecules, so that there can be fluctuation in the dipole-dipole interaction~(DDI). Because this fluctuation hampers the gate fidelity~\cite{Ni_2018,bergonzoni_iswap_2025}, the molecules were usually cooled to near the ground QM state~\cite{ruttley_long-lived_2025}. Even in the motional ground state, the QM is still there.

In this work, we unveil two novel consequences of the QM when treating the in-trap molecular motion in a quantum mechanical framework: (1) It can realize the asymmetric quantum Rabi model~(AQRM)~\cite{Rabi_1936,Rabi_1937,Braak_2011}, where the two DDI-coupled two-molecule internal states form the two-level matter system in the standard quantum Rabi model~(QRM), while the bosonic motional modes are analogous to the photonic modes in the widely studied QRM based on cavity QED. This makes polar molecules an alternative platform for simulating QRM which is of great interest in the study of fundamental physics as well as the exploration of strong and ultrastrong couplings between light and matter~\cite{Forn_D_az_2019,Frisk_Kockum_2019}; (2) The molecular QM can lead to a trap-dipole resonance, which is due to a resonant energy exchange between the QM of the real-space in-trap motion and the internal DDI. We find that this resonance occurs at $J_0/(\hbar\omega)\approx1$ and 2, with $\omega$ one of the trap frequencies and $J_0$ the magnitude of the DDI. This resonance leads to excess population loss, and, hence, should be avoided in a general control with polar molecules.

To understand the impact of QM-DDI coupling on quantum information processing with polar molecules~\cite{DeMille_2002,Zhu_2013,Yelin_2006,Ni_2018,Hughes_2020,bergonzoni_iswap_2025} where entangling gate can be generated by DDI~\cite{picard_entanglement_2025}, we introduce two gate protocols: (1) A modified iSWAP gate which can work with $J_0/(\hbar\Omega_\mu)\sim1/2$ and can be rapidly realized with one microwave pulse of pulse area less than $2\pi$, with $\Omega_\mu$ the Rabi frequency and $\hbar$ the reduced Planck constant. This is in contrast to the standard iSWAP gate which needs a wait duration between two microwave $\pi$ pulses~\cite{picard_entanglement_2025} and where the gate fidelity is high when $J_0/(\hbar\Omega_\mu)$ approaches zero. (2) A controlled-phase gate with an arbitrarily desired phase which depends on the blockade effect but can attain a high fidelity with $J_0/(\hbar\Omega_\mu)$ around ten. Both gates introduced in this work can attain a high fidelity in the presence of QM in typical setups.

The remainder of this article is as follows. In Sec.~\ref{sec-rabi}, we introduce the AQRM based on QM of polar molecules. In Sec.~\ref{sec-resoance}, we study the trap-dipole resonance. In Sec.~\ref{sec-swap}, we present an iSWAP gate implemented by one microwave pulse. In Sec.~\ref{sec-quasi}, we introduce a quasi-blockade two-qubit controlled phase gate. Both Sec.~\ref{sec-swap} and Sec.~\ref{sec-quasi} show results of the gate fidelity in the presence of QM-DDI coupling. We give discussions in Sec.~\ref{sec-06} and summarize in Sec.~\ref{sec-07}.

\begin{figure}
\includegraphics[width=3.0in]
{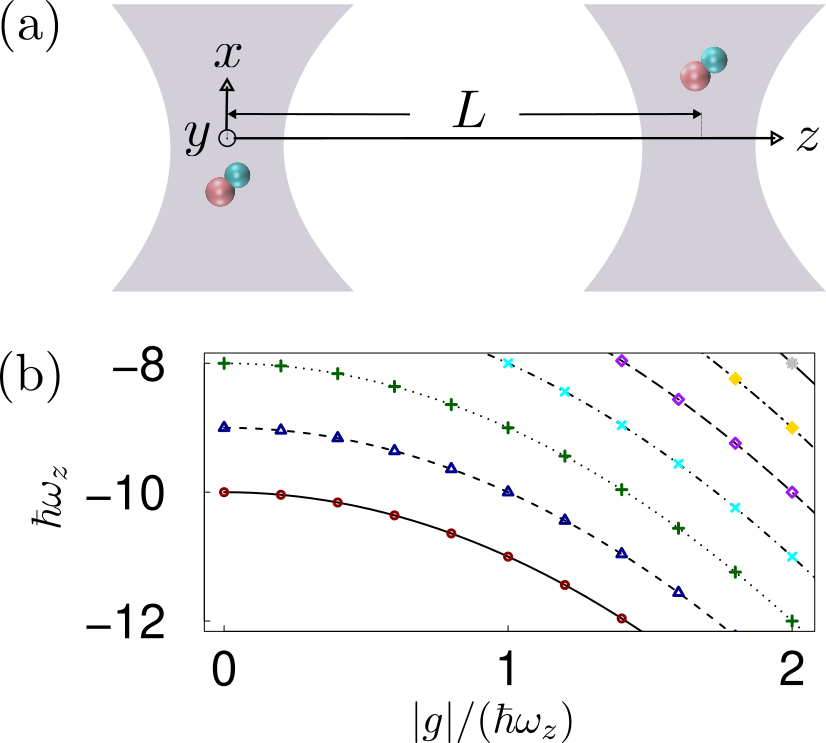}
\caption{(a) Illustration of two optically trapped polar molecules in two traps separated by $L$ along z. Due to the QM, the actual positions of the molecules are not necessarily at the centers of the traps. (b) The black curves show part of the numerically calculated eigenspectrum of the AQRM Hamiltonian in Eq.~(\ref{Rabi-01}) as a function of $g=-3J_0\ell_z/L$ when $J_0=10\hbar\omega_z$. Motional quanta up to 60 are included in the numerical codes. The colored symbols show eigenspectrum with the analytical eigenstates given in Eq.~(\ref{Rabi-04}). One can see that the numerical results match well with the analytical results.   \label{figure-1d} }
\end{figure}
\section{Asymmetric quantum Rabi model with polar molecules}~\label{sec-rabi}
In this section, we describe the details of QM-DDI coupling in two polar molecules trapped in two nearby optical traps, and show how the coupling leads to an asymmetric QRM.

\subsection{QM-DDI coupling}\label{sec-rabi-coupling}
Consider two molecules, labeled c and t, where c and t are abbreviations of ``control'' and ``target'' to be compatible with the study of quantum gates in Secs.~\ref{sec-swap} and~\ref{sec-quasi}. The molecules are trapped in optical traps centered at $(0,0,0)$ and $(0, 0,L)$, respectively, as illustrated in Fig.~\ref{figure-1d}(a). Though the presentation focuses on the case of optical tweezers, the results in this work are applicable for optical lattices. The Hamiltonian for the two trapped molecules is
\begin{eqnarray}
 \hat{H}_{\text{trap}} &=& \sum_{\alpha=\text{c,t}} \sum_{\xi =\text{x,y,z}}\left\{ -\frac{\hbar^2}{2m} \frac{\partial^2}{\partial \xi_\alpha^2 }  +\frac{m\omega_\xi^2\xi_\alpha^2}{2}\right\}\nonumber\\  &&+ \frac{m\omega_z^2 }{2}\left[(z_{\text{t}}-L)^2- z_{\text{t}}^2 \right]  ,\label{H-0}
\end{eqnarray}
where $\hbar$ $m$ is the mass of the molecule, and $\omega_\xi$ is the angular trap frequency along $\xi\in\{x,y,z\}$. To capture the coupling between the motion and the internal states of the trapped molecules, we define $z_\pm = (z_{\text{c}}\pm z_{\text{t}} \mp L )/\sqrt{2} $ for the z component~\cite{Lu2026gate}, and
\begin{eqnarray}
 \xi_{\pm}&=&\frac{1}{\sqrt{2}}(\xi_{\text{c}}\pm \xi_{\text{t}}) \label{transform}
\end{eqnarray}
for the $\xi =x$ and $y$ component, based on which one finds that Eq.~(\ref{H-0}) becomes,
\begin{eqnarray}
 \hat{H}_{\text{trap}} &=& \sum_{\alpha=\pm} \sum_{\xi =\text{x,y,z}}\left\{ -\frac{\hbar^2}{2m} \frac{\partial^2}{\partial \xi_\alpha^2 }  +\frac{m\omega_\xi^2\xi_\alpha^2}{2}\right\}  ,\label{H-1}
\end{eqnarray}
which means that the state of the two-molecule vibration in the two-trap system is a product state of the two motional modes labeled $\pm$. The harmonic oscillator lengths are $\ell_\xi =  \sqrt{ \hbar/(m\omega_\xi)}$ along the three directions. We define
\begin{eqnarray}
\hat{a}_{\xi} &=&  \frac{1}{\sqrt{2 } \ell_\xi}\left(\xi_-+ \ell_\xi^2 \frac{\partial}{\partial \xi_-}\right),
\hat{a}_{\xi}^\dag=  \frac{1}{\sqrt{2 } \ell_\xi}\left(\xi_-- \ell_\xi^2 \frac{\partial}{\partial \xi_-}\right),\nonumber\\
\hat{b}_{\xi} &=&  \frac{1}{\sqrt{2  } \ell_\xi}\left(\xi_+ + \ell_\xi^2 \frac{\partial}{\partial \xi_+}\right),
\hat{b}_{\xi}^\dag =  \frac{1}{\sqrt{2  } \ell_\xi}\left(\xi_+ - \ell_\xi^2 \frac{\partial}{\partial \xi_+}\right),\nonumber\\\label{a-b-mode}
\end{eqnarray}
so that Eq.~(\ref{H-1}) can be further written as
\begin{eqnarray}
 \hat{H}_{\text{trap}} &=&  \hbar \sum_{\xi =\text{x,y,z}}\omega_\xi(\hat{a}_{\xi }^\dag \hat{a}_{\xi} +\hat{b}_{\xi }^\dag \hat{b}_{\xi} +1),\label{H-trap01}
\end{eqnarray}
where $\{\hat{a}_{\xi }^\dag ,\hat{a}_{\xi}\}$ and $\{\hat{b}_{\xi }^\dag ,\hat{b}_{\xi}\}$ are the bosonic creation and annihilation operators for the motional states of the two-molecule motional mode $\alpha=$- and + along $\xi$, respectively. They are related to $\xi_\alpha$ via $\xi_- = \frac{\ell_\xi}{\sqrt{2}}(\hat{a}_{\xi}^\dag + \hat{a}_{\xi}) $ and $\xi_+ = \frac{\ell_\xi}{\sqrt{2}}(\hat{b}_{\xi}^\dag + \hat{b}_{\xi}) $.

We consider two different molecular states labelled $\lvert\uparrow\rangle$ and $|e\rangle$, and a condition that a two-molecular state in a superposition of $\lvert\uparrow,e\rangle$ of $\lvert e,\uparrow\rangle$ experiences a DDI~\cite{Ni_2018,picard_entanglement_2025}, where $\lvert\uparrow,e\rangle\equiv \lvert \uparrow_{\text{c}}\rangle\otimes\lvert e_{\text{t}}\rangle$ and similar for $\lvert e,\uparrow\rangle$. Following Refs.~\cite{ruttley_long-lived_2025,picard_entanglement_2025}, we consider that the quantization axis is along the two-tweezers separation $\mathbf{z}$. The DDI between two electric dipole moments is~\cite{Shi2021qst},
\begin{eqnarray}
\hat{V}&=&\frac{1 }{ 4\pi\epsilon_0 r^3 }   \hat{\mathbf{s}}_{\text{c}} \cdot\left(3\frac{\hat{\mathbf{r}} \hat{\mathbf{r}}}{r^2} -\hat{\mathbf{I}} \right) \cdot\hat{\mathbf{s}}_{\text{t}} \label{V-operator}
\end{eqnarray}
where $\epsilon_0$ is the vacuum permittivity, $\hat{\mathbf{r}}\equiv \hat{\mathbf{r}}_{\text{c}}-\hat{\mathbf{r}}_{\text{t}}$ is the separation vector between the two molecules at $\hat{\mathbf{r}}_{\text{c}}$ and $\hat{\mathbf{r}}_{\text{t}}$, respectively, with $r$ its magnitude, and $\hat{\mathbf{s}}_\alpha$ is the electric dipole operator. Even in the restriction of selection rules and that DDI does not change the nuclear spin states, Eq.~(\ref{V-operator}) shows that there can be many states coupled. However, typical polar molecules near the rovibrational ground manifold have certain hyperfine-Zeeman substates that are coupled only to the state-flipped counterpart, while all the other states are far off-resonant, resulting in a nearly perfect spin-exchange interaction~\cite{Ni_2018}. In other words, we can have $\hat{V} = J \left(\lvert \uparrow,e\rangle\langle e,\uparrow\rvert+ \lvert e,\uparrow\rangle\langle \uparrow,e\rvert\right)$ by projecting Eq.~(\ref{V-operator}) onto the state space connected by the dipole-dipole interaction
\begin{eqnarray}
J &=&J_0  \frac{1-3\cos^2\theta}{2} \frac{L^3}{|\hat{\mathbf{r}}_{\text{c}}-\hat{\mathbf{r}}_{\text{t}}|^3},\label{J-fluctuation0-0}
\end{eqnarray}
where $J_0 =\frac{1 }{ 2\pi\epsilon_0L^3}  \left(\frac{\text{\textdong}}{\sqrt{3}} \right)^2$, with \textdong~the molecule-frame electric dipole moment, and $\theta$ is the angle between the quantization axis and the separation axis of the two molecules. In this work, the DDI is denoted sometimes by $J$ when no QM-DDI coupling is considered, and sometimes by $J_0$ which denotes the magnitude of DDI when molecules are exactly in the centers of the trap.

We further define $\epsilon_\xi= \sqrt{2}\xi_-/L$ with $\xi_-$ given in Eq.~(\ref{transform}). Then, Eq.~(\ref{J-fluctuation0-0}) can be written as
\begin{eqnarray}
J &=& -\frac{J_0}{2}\frac{2(1-\epsilon_z )^2 - (\epsilon_x^2+\epsilon_y^2 ) }{[\epsilon_x^2+\epsilon_y^2 +(1-\epsilon_z )^2]^{5/2} ×}. \label{J-fluctuation0}
\end{eqnarray}

According to the relation between the position operator and $\hat{a}$ shown below Eq.~(\ref{H-trap01}), Eq.~(\ref{J-fluctuation0}) is equivalent to
\begin{eqnarray}
J &=& -\frac{J_0}{2}\bigg\{ 2[( \hat{a}_{z}^\dag +\hat{a}_{z})\ell_z -L ]^2 - \ell_x^2(\hat{a}_{x}^\dag +\hat{a}_{x})^2\nonumber\\
&& - \ell_y^2(\hat{a}_{y}^\dag +\hat{a}_{y})^2 \bigg\} \bigg\{  [( \hat{a}_{z}^\dag +\hat{a}_{z})\ell_z -L ]^2\nonumber\\
&& +  \ell_x^2(\hat{a}_{x}^\dag +\hat{a}_{x})^2 +  \ell_y^2(\hat{a}_{y}^\dag +\hat{a}_{y})^2 \bigg\}^{-5/2}. \label{J-fluctuation0-1}
\end{eqnarray}
Equation~(\ref{J-fluctuation0-1}) shows that the QM-DDI coupling involves three sets of bosonic modes, i.e., the modes represented by $\{\hat{a}_\xi,~\hat{a}_\xi^\dag\}$ with $\xi\in\{x,y,z\}$, while the other three set of modes $\{\hat{b}_\xi,~\hat{b}_\xi^\dag\}$ are not coupled. As a result, the three sets of modes $\{\hat{b}_\xi,~\hat{b}_\xi^\dag\}$ with $\xi\in\{x,y,z\}$ are not involved in DDI. As long as DDI is concerned, we can ignore the three set of idling modes, so that Eq.~(\ref{H-trap01}) can be written as
\begin{eqnarray}
 \hat{H}_{\text{trap}} &=&  \hbar \sum_{\xi =\text{x,y,z}}\omega_\xi(\hat{a}_{\xi }^\dag \hat{a}_{\xi}   +1/2),\label{H-trap01-2}
\end{eqnarray}
where the term $\hbar\omega_\xi/2$ is the ground-state energy for the mode along $\xi$.

\subsection{Asymmetric quantum Rabi model without the $\sigma_3$ term}\label{sec-rabi-A}
We consider conditions when $|\epsilon_\xi|\ll 1$ for all $\xi=x,y$, and $z$, so that we can keep terms up to square of these small parameters, leading to
\begin{eqnarray}
J &=& - J_0[1+3\epsilon_z+6\epsilon_z^2 -3(\epsilon_x^2+\epsilon_y^2)]. \label{J-fluctuation2}
\end{eqnarray}
According to the relation between the position operator and the bosonic operators $\hat{a}_\xi$, Eq.~(\ref{J-fluctuation2}) can be further written as
\begin{eqnarray}
J &=& - \frac{J_0}{L^2}\left\{L^2+3\ell_zL(\hat{a}_{z}^\dag +\hat{a}_{z} )+ 6\ell_z^2(\hat{a}_{z}^\dag +\hat{a}_{z})^2 \right.\nonumber\\
&&\left.-3[\ell_x^2(\hat{a}_{x}^\dag +\hat{a}_{x})^2 +\ell_y^2(\hat{a}_{y}^\dag +\hat{a}_{y})^2]\right\}. \label{J-fluctuation3}
\end{eqnarray}

To show that the trap-molecule system can realize the simplest version of the AQRM, we consider a condition when $|\ell_z/L|\gg \ell_x^2/L^2,\ell_y^2/L^2$~\footnote{This condition differs from current experiments in, e.g., Refs.~\cite{ruttley_long-lived_2025,picard_entanglement_2025}: in reference to the setup of Fig.~\ref{figure-1d}, $\ell_z/L$ is comparable to the axial $(\ell_x/L)^2$ in Refs.~\cite{ruttley_long-lived_2025}, and in Ref.~\cite{picard_entanglement_2025} the only relevant motion is along x. }. Further, because $|\ell_\xi/L|$ in Eq.~(\ref{J-fluctuation3}) are small for all $\xi\in\{x,y,z\}$, the term with $\ell_z^2/L^2$ is small compared to the term with $\ell_z/L$. In this case, we have $J\approx  -  J_0 [1+3\ell_z(\hat{a}_{z}^\dag +\hat{a}_{z} )/L]$. Then, the Hamiltonian $\hat{H} = \hat{H}_{\text{trap}}   +\hat{V}\approx\hbar\omega_z\hat{a}_{z}^\dag \hat{a}_{z} +\hat{V}$ for the molecular motion and DDI can be written as
\begin{eqnarray}
\hat{H} &=& \hbar\omega_z\hat{a}_{z}^\dag \hat{a}_{z} +g\sigma_1( \hat{a}_{z}^\dag +\hat{a}_{z}) + \eta\sigma_1,~\label{Rabi-01}
\end{eqnarray}
with $(\eta,~g) =-J_0(1,~3\frac{\ell_z}{L}) $ and $\sigma_1= \lvert \uparrow,e\rangle\langle e,\uparrow\rvert+ \lvert e,\uparrow\rangle\langle \uparrow,e\rvert$, where we have dropped off a constant $\hbar\omega_z/2$ for it only induces a global linear phase shift. Equation~(\ref{Rabi-01}) is a Hamiltonian of the AQRM~\cite{Braak_2011}. Both the QRM and the AQRM are of specific interest in the exploration of the ultrastrong coupling between light and matter~\cite{Forn_D_az_2019,Frisk_Kockum_2019}. We note that for QRM, there are some proposals for realizing it with atomic and molecular systems~\cite{Felicetti_2017,Hunanyan_2024}.

Equation~(\ref{Rabi-01}) can be rewritten as
\begin{eqnarray}
\hat{H} &=& \hbar\omega_z\hat{a}_{z}^\dag \hat{a}_{z} +[g ( \hat{a}_{z}^\dag +\hat{a}_{z}) + \eta](\lvert+\rangle\langle +\rvert -\lvert-\rangle\langle -\rvert ),\nonumber\\~\label{Rabi-02}
\end{eqnarray}
where $\vert\pm\rangle =(\lvert \uparrow,e\rangle \pm \lvert e,\uparrow\rangle)/\sqrt{2}$. We use a displacement operator $\Phi^{(\pm)}=$exp$[\pm(\hat{a}_z^\dag-\hat{a}_z)g/(\hbar\omega_z)]$ which can generate the following operators via $\Phi^{(\pm)\dag} (\cdots)\Phi^{(\pm)}$,
\begin{eqnarray}
\hat{A}_z &=& \hat{a}_z+ g/(\hbar\omega_z),~\hat{A}_z^{\dag} = \hat{a}_z^\dag+ g/(\hbar\omega_z),\nonumber\\
\hat{A}_{\underline{z}}&=& \hat{a}_z- g/(\hbar\omega_z),~\hat{A}_{\underline{z}}^\dag = \hat{a}_z^\dag- g/(\hbar\omega_z),\label{z-tran-01}
\end{eqnarray}
so that Eq.~(\ref{Rabi-02}) becomes
\begin{eqnarray}
\hat{H} &=& [\hbar\omega_z\hat{A}_{z}^\dag \hat{A}_{z} -g^2/(\hbar\omega_z)+ \eta]\lvert+\rangle\langle +\rvert  \nonumber\\
&&+ [\hbar\omega_z\hat{A}_{\underline{z}}^\dag \hat{A}_{\underline{z}} -g^2/(\hbar\omega_z)- \eta] \lvert-\rangle\langle -\rvert ),~\label{Rabi-03}
\end{eqnarray}
which means that the eigenstates of the AQRM of Eq.~(\ref{Rabi-01})
are
\begin{eqnarray}
\lvert N_z \rangle^{(\pm)}&=&\frac{1}{\sqrt{N_z!}} \lvert\pm\rangle\otimes (\hat{A}_{z(\underline{z})}^\dag)^{N_z}\lvert\text{vac}\rangle,~\label{Rabi-04}
\end{eqnarray}
of eigenenergy $\hbar\omega_zN_z -g^2/(\hbar\omega_z)\pm \eta$, which is shown by colored symbols in Fig.~\ref{figure-1d}, and a general state of the model is $\alpha\lvert N_z\rangle^{(+)}+\beta\lvert N_{z'}\rangle^{(-)}$ with $|\alpha^2|+|\beta|^2=1$. To check if the above procedure is correct, we also use numerical diagonization to calculate the eigenenergy of the Hamiltonian~(\ref{Rabi-01}), with results given by the black curves. One can see that the analytical results match well with the numerical results.

\begin{figure}
\includegraphics[width=3.5in]
{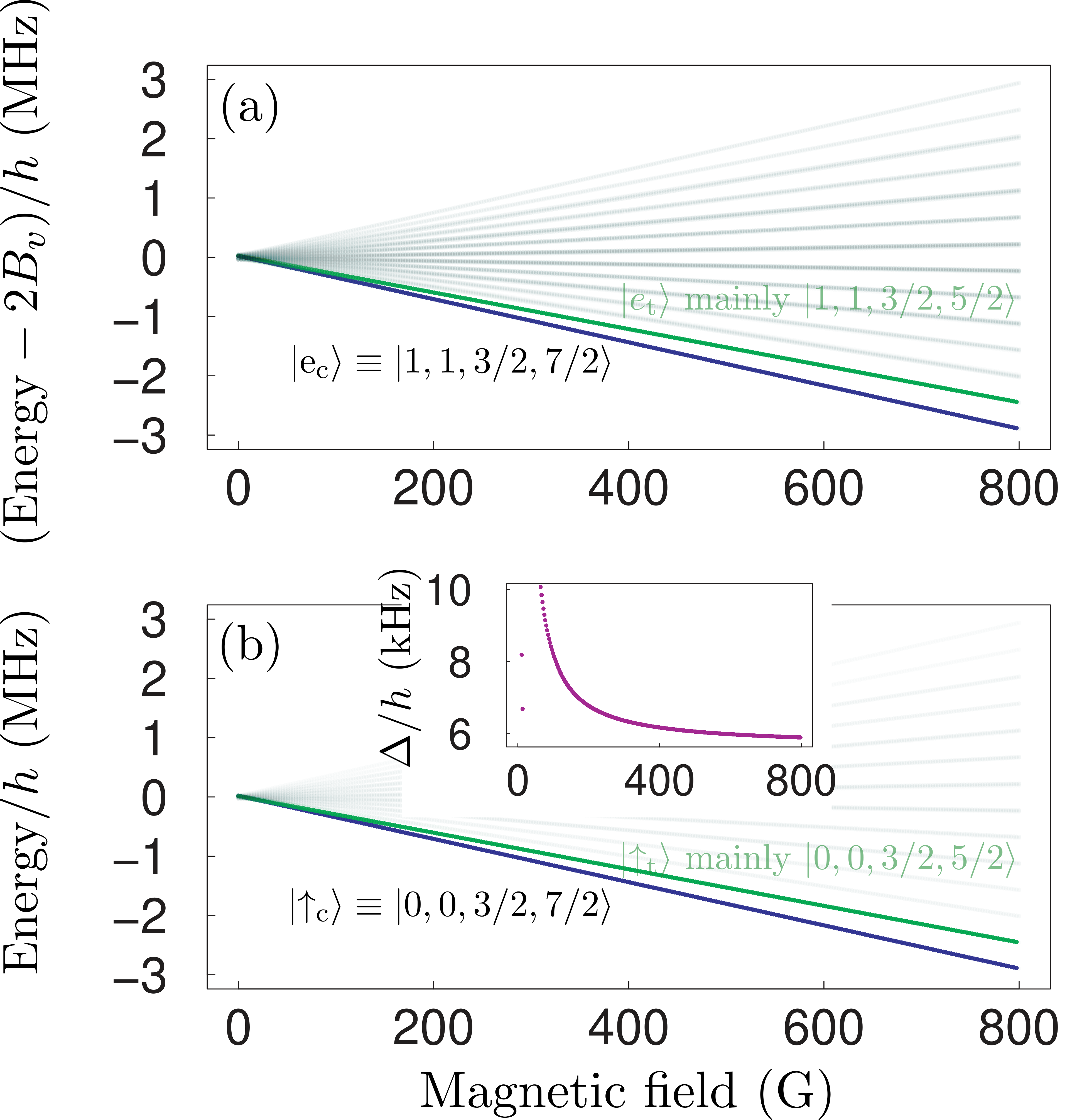}
\caption{Feasibility to realize a nonzero $\Delta$ in Eq.~(\ref{Rabi-02-sigma3}).  (a,b) shows the energy of the $N=1$ and $N=0$ rovibrational ground manifold of $^{23}$Na$^{133}$Cs in zero electric field, where the energy in (a) is shifted by $2B_v$ with $B_v$ the rotational constant of the rovibrational ground manifold.     \label{figure-hybrid} }
\end{figure}

The model in Eq.~(\ref{Rabi-01}) does not have the term $\Delta\sigma_3$, where $\sigma_3\equiv \lvert \uparrow,e\rangle\langle \uparrow,e\rvert - \lvert e, \uparrow\rangle\langle e,\uparrow\rvert$. In the context of the standard QRM, $\Delta$ is the energy separation between the two levels in the two-level matter qubit. To our knowledge, there is no counterpart for neither QRM nor AQRM with $\Delta=0$, which means that the model in this article enables the exploration of AQRM in a new domain.

\subsection{AQRM with the $\sigma_3$ term}\label{sec-rabi-B}
The QM-DDI coupling of polar molecules not only can emulate the exotic AQRM without the $\sigma_3$ term, but can also simulate the standard AQRM with the term $\Delta\sigma_3$. To achieve this, we can choose different DDI-coupled pair states for the two molecules. Take the $^1\Sigma^+$ ground manifold of $^{23}$Na$^{133}$Cs as an example, we can choose
\begin{eqnarray}
\lvert\uparrow_{\text{c}}\rangle &\equiv& \lvert0,0,3/2,7/2\rangle,\nonumber\\
\lvert e_{\text{c}}\rangle &\equiv& \lvert1,1,3/2,7/2\rangle,
\end{eqnarray}
for molecule c. Whether the B-field is weak or strong, there is no hyperfine-induced state mixing because their spin projections are maximally stretched, which is a usual character in hyperfine interactions of atomic and molecular systems~\cite{PhysRevA.107.023102}. For molecule t, we can choose, for example,
\begin{eqnarray}
\lvert\uparrow_{\text{t}}\rangle &\equiv~(\text{mainly has})& \lvert0,0,3/2,5/2\rangle,\nonumber\\
\lvert e_{\text{t}}\rangle &\equiv~(\text{mainly has})& \lvert1,1,3/2,5/2\rangle,
\end{eqnarray}
where the right sides of the above equations show the state components with the largest portion. Due to the hyperfine interaction and interaction involving the nuclear spins, $\lvert\uparrow_{\text{t}}\rangle$ and $\lvert e_{\text{t}}\rangle$ are composed of two and three hyperfine-Zeeman substates, respectively
\begin{eqnarray}
\lvert\uparrow_{\text{t}}\rangle &=& \alpha_1\lvert0,0,3/2,5/2\rangle+\alpha_2\lvert0,0,1/2,7/2\rangle,\nonumber\\
\lvert e_{\text{t}}\rangle &\equiv& \beta_1\lvert1,1,3/2,5/2\rangle +  \beta_2\lvert1,1,1/2,7/2\rangle\nonumber\\
&& + \beta_3\lvert1,0,3/2,7/2\rangle .
\end{eqnarray}
Near zero magnetic field, the values of $|\alpha_1|^2 $ and $|\beta_1|^2$ are about 0.7 and 0.58, and the population of $\lvert0,0,1/2,7/2\rangle$ and $\lvert1,1,1/2,7/2\rangle$ in $\lvert\uparrow_{\text{t}}\rangle$ and $\lvert e_{\text{t}}\rangle$ are about 0.3 and 0.25. Except of populating $\lvert1,1,3/2,5/2\rangle$ and $\lvert1,1,1/2,7/2\rangle$, there is some population in $\lvert1,0,3/2,7/2\rangle$ for the state $\lvert e_{\text{t}}\rangle$, about 0.17. This last state will cause issue because for the state at $\lvert \uparrow_{\text{c}}, e_{\text{t}}\rangle$, though the major component $\lvert1,1,3/2,5/2\rangle$ in molecule t cannot go to the state $\lvert0,0,3/2,7/2\rangle$, the component $\lvert1,0,3/2,7/2\rangle$ can. In this case, there will still be resonant interaction.

To suppress the resonant DDI, we can apply a strong magnetic field. This is because the above issue caused by hyperfine or nuclear spin involved interactions can be suppressed by strong magnetic field, as shown in Fig.~\ref{figure-hybrid} which was calculated via Diatomic-py~\cite{Blackmore_2023}. At $B_z=800$~G, the values of $|\alpha_1|^2 $ and $|\beta_1|^2$ are about 0.9996 and 0.9990, and the population in $\lvert1,0,3/2,7/2\rangle$ for the state $\lvert e_{\text{t}}\rangle$ is only $6.1\times10^{-4}$. In this case, for the initial two-molecule state $\lvert \uparrow_{\text{c}}, e_{\text{t}}\rangle$, the DDI will couple it to $\lvert e_{\text{c}},\uparrow_{\text{t}}\rangle$, which is lower by $\Delta$. The value of $\Delta$ as a function of the magnetic field is shown in the inset of Fig.~\ref{figure-hybrid}(b). In Fig.~\ref{figure-hybrid}(a,b), the $N=1$ and $N=0$ part of the Zeeman spectra of the molecule is shown, where the blue~(green) color denotes the states for molecule c~(t).

The example of Fig.~\ref{figure-hybrid} shows that at $B_z=800$~G, $\Delta/h$ is about
$5.9$~kHz. If one would like to have a smaller $\Delta$, then other states can be used for molecule t. For example, if we choose
\begin{eqnarray}
\lvert\uparrow_{\text{t}}\rangle &\equiv~(\text{mainly has})& \lvert0,0,1/2,-1/2\rangle,\nonumber\\
\lvert e_{\text{t}}\rangle &\equiv~(\text{mainly has})& \lvert1,1,1/2,-1/2\rangle,
\end{eqnarray}
then $\Delta/h$ is about $0.44$~(0.15)~kHz at $B_z=800~(2000)$~G, while the population of $\lvert0,0,1/2,-1/2\rangle$ and $\lvert1,1,1/2,-1/2\rangle$ in $\lvert\uparrow_{\text{t}}\rangle$ and $\lvert e_{\text{t}}\rangle$ are about 0.9980 and 0.9967~(0.9997 and 0.9996), respectively, large enough for the desired DDI coupling to occur. If smaller $\Delta$ is needed, large B-field can be applied. For example, if $B_z= 5000$~G, $\Delta/h$ is only 0.034~kHz. As a result, we can have the AQRM with the $\sigma_3$ term, i.e., Eq.~(\ref{Rabi-01}) is updated to
\begin{eqnarray}
\hat{H} &=& \hbar\omega_z\hat{a}_{z}^\dag \hat{a}_{z} +g\sigma_1( \hat{a}_{z}^\dag +\hat{a}_{z}) + \eta\sigma_1 +  \Delta\sigma_3/2,~\label{Rabi-02-sigma3}
\end{eqnarray}
with a trivial constant term $-\Delta/2$ ignored.

\subsection{ 3D asymmetric quantum Rabi model}~\label{sec-rabi3}
It is also possible to realize a three dimensional~(3D) AQRM, which is of particular relevance to recent experiments. Consider $(\omega_x,\omega_y,\omega_z)=2\pi\times(0.4,~3.0,~3.0)$~kHz as like the experiment of Ref.~\cite{ruttley_long-lived_2025} where $^{87}$Rb$^{133}$Cs molecules were trapped in tweezer arrays with spacing $L=2.78~\mu$m, resulting in $(\ell_x,~\ell_y,~\ell_z)/L\approx (0.12,~0.045,~0.045)$. In this case, it is desirable to preserve the three modes in Eq.~(\ref{J-fluctuation3}). Then,  $ \hat{H}_{\text{trap}}   +\hat{V}$ can be written as
\begin{eqnarray}
\hat{H} &=& \sum_\xi   [ \hbar\omega_{\xi}\hat{a}_{\xi}^\dag \hat{a}_{\xi} + \hat{h}_\xi \sigma_1]+\eta\sigma_1,\label{3dAQRM}
 \end{eqnarray}
  where
\begin{eqnarray}
\hat{h}_x  &=& \zeta_x(\hat{a}_x^\dag +\hat{a}_x)^2 ,\nonumber\\
\hat{h}_y &=&\zeta_y(\hat{a}_y^\dag +\hat{a}_y)^2  ,\nonumber\\
\hat{h}_z &=& g(\hat{a}_z^\dag +\hat{a}_z)+\zeta_z(\hat{a}_z^\dag +\hat{a}_z)^2 ,\nonumber
\end{eqnarray}
and
\begin{eqnarray}
 (\zeta_x,~\zeta_y,~\zeta_z) &=& 3\frac{J_0  }{  L^2}(\ell_x^2,~\ell_y^2,~-2\ell_z^2).
 \end{eqnarray}
By comparing Eq.~(\ref{3dAQRM}) and Eq.~(\ref{Rabi-01}), one can see that we have a 3D AQRM with Eq.~(\ref{3dAQRM}).

There is no direct coupling between the three motional modes, but they are indirectly coupled via the two-level system. To show this, one can see that Eq.~(\ref{3dAQRM}) can be written as
\begin{eqnarray}
\hat{H} &=& \sum_\xi \sum_{\alpha=\pm} [ \hat{a}_{\xi}^\dag \hat{a}_{\xi}\hbar\omega_{\xi} +\alpha\hat{h}_\xi+\alpha\eta/3 ]\lvert \alpha\rangle\langle \alpha\rvert,\label{3dAQRM-2}
 \end{eqnarray}
with $\lvert\pm\rangle$ given below Eq.~(\ref{Rabi-02}). We introduce a Bogoliubov transform~\cite{Bogo1947} for each mode,
\begin{eqnarray}
\mathscr{B}_\xi^{(\pm)}&=&\text{exp}\{{\varpi}_{\xi}^{(\pm)}[(\hat{a}_{\xi}^\dag )^2 - \hat{a}_\xi^2]\},\nonumber\\
{\varpi}_\xi^{(\pm)} &=&\frac{1}{8}\ln \left(1\pm \frac{4\zeta_\xi}{\hbar\omega_\xi}\right)\label{Bogn01}
 \end{eqnarray}
 in the regime of $\frac{\ell_\xi^2}{L^2}<\frac{\hbar\omega_\xi}{12J_0}$ when $\xi=x$ or $y$ and $\frac{\ell_z^2}{L^2}<\frac{\hbar\omega_z}{24J_0}$, and further a displacement transform for the $z$ mode via
\begin{eqnarray}
\Phi^{(\pm)}&=&\text{exp}\{\pm[ \hat{a}_\xi^\dag-   \hat{a}_{\xi} ]g e^{-6\varpi_{z}^{(\pm)}}/(\hbar\omega_z)\},\label{Bogn02}
 \end{eqnarray}
which generates a new sets of operators for the squeezed modes,
\begin{eqnarray}
\hat{A}_{\xi(\overline{\xi})} &=&\mathscr{B}_\xi^{(\pm)\dag} \hat{a}_{\xi(\overline{\xi})}\mathscr{B}_\xi^{(\pm)},~\text{when }\xi\in\{x,~y\},\nonumber\\
\hat{A}_{z(\overline{z})} &=&\Phi^{(\pm)\dag}\mathscr{B}_z^{(\pm)\dag} \hat{a}_{z(\overline{z})}\mathscr{B}_z^{(\pm)}\Phi^{(\pm)}.
 \end{eqnarray}
Insertion of the new operators into Eq.~(\ref{3dAQRM-2}) leads to
\begin{eqnarray}
\hat{H} &=& \sum_\xi  \left [ \left(\hat{A}_{\xi}^\dag \hat{A}_{\xi}+\frac{1}{2}\right) \hbar\omega_\xi e^{4\varpi_\xi^{(+)}} - \frac{1}{2}) \hbar\omega_\xi \right]\lvert +\rangle\langle + \rvert\nonumber\\ && + \sum_\xi  \left [ \left(\hat{A}_{\overline{\xi}}^\dag \hat{A}_{\overline{\xi}}+\frac{1}{2}\right) \hbar\omega_\xi e^{4\varpi_{\xi}^{(-)}} - \frac{1}{2}) \hbar\omega_\xi \right]\lvert -\rangle\langle - \rvert\nonumber\\&& + \sum_{\alpha=\pm}\left[ \alpha\eta - \frac{g^2}{\hbar\omega_z\pm 4\zeta_\xi} \right]\lvert \alpha\rangle\langle \alpha \rvert.
 \end{eqnarray}
 The eigenstate for the system is $\lvert N_x,N_y,N_z \rangle^{(\pm)}=\frac{1}{\sqrt{N_x!N_y!N_z!}} \lvert\pm\rangle\otimes [\hat{A}_{x(\underline{x})}^\dag]^{N_x} [\hat{A}_{y(\underline{y})}^\dag]^{N_y}[\hat{A}_{z(\underline{z})}^\dag]^{N_z}\lvert\text{vac}\rangle$.

\section{ Trap-dipole resonance}\label{sec-resoance}
DDI of optically trapped single molecules can be used for entanglement generation~\cite{DeMille_2002,Zhu_2013,Yelin_2006,Ni_2018,Hughes_2020,Tscherbul_2023,bergonzoni_iswap_2025,Muminov_2026}, which has been recently realized in experiments~\cite{Bao_2023,Holland_2023,ruttley_long-lived_2025,picard_entanglement_2025}. Here, we show that there will be a detrimental trap-dipole resonance which should be avoided.

To be consistent with a gate we will introduce later, we consider three states, two qubit states $\lvert \uparrow\rangle,~\lvert \downarrow\rangle$, and an excited state $\lvert e\rangle$ as in Ref.~\cite{picard_entanglement_2025}, where there is DDI with two molecules in states of $\lvert \uparrow,e\rangle $ and $\lvert e,\uparrow\rangle$. Here, we consider the excitation blockade effect which is useful for molecular entanglement~\cite{ruttley_long-lived_2025} or quantum gate~\cite{Yelin_2006}. The blockade effect can work in various ways with the same physical principle~\cite{Shi2021qst}, i.e., the Rabi frequency $\Omega_\mu$ being much smaller than $J/\hbar$ in our case. With the Hamiltonian of the microwave field 
\begin{eqnarray}
\hat{H}_{\mu} = \hbar\frac{\Omega_{\mu}(t)}{2}\lvert e\rangle\langle\downarrow\rvert +\text{H.c.} ,\label{H_mu-1}
\end{eqnarray}
the two-molecular Hamiltonian is $ \hat{H}= \hat{H}_{\mu}\otimes\hat{I}+\hat{I}\otimes\hat{H}_{\mu}+\hat{V}$ with $\hat{I}$ the identity operator. Because each of the two molecules with the input state $\lvert\downarrow \downarrow\rangle$ will have a Rabi oscillation with the excited state $\lvert e\rangle$, which is easily understood, we only focus on the time dynamics for the input states $\lvert \uparrow \downarrow\rangle$ and $\lvert \downarrow \uparrow \rangle$, whose Hamiltonian can be written as $\hat{H}=\hat{H}_++ \hat{H}_-$, where
\begin{eqnarray}
\hat{H}_\pm &=&\left[\hbar\frac{\Omega_{\mu}(t)}{2}\lvert \pm\rangle\langle\mathbb{B}_\pm\rvert +\text{H.c.}\right]\pm J\lvert \pm\rangle\langle\pm \rvert, \label{H_iswap}
\end{eqnarray}
where $\lvert \mathbb{B}_\pm\rangle =\frac{1}{\sqrt{2}}\left( \lvert \uparrow \downarrow\rangle\pm  \lvert \downarrow \uparrow \rangle \right)$.
When $\hbar|\Omega_\mu/J|\ll 1$ in the blockade condition~\cite{ruttley_long-lived_2025,Yelin_2006}, a $\pi$ pulse will induce $\lvert\downarrow \downarrow\rangle\rightarrow-\lvert ee\rangle$ but nothing occurs for $ \lvert\mathbb{B}_+\rangle,\lvert \mathbb{B}_-\rangle$. Then applying a second $\pi$ pulse but with $i\Omega_\mu$, i.e., a $\pi/2$ phase change to the microwave field, then $ -\lvert ee\rangle\rightarrow-\lvert\downarrow \downarrow\rangle$, so that we have a gate diag$\{1,1,1,-1\}$ in the basis of $\{\lvert\uparrow \uparrow\rangle,  \lvert\mathbb{B}_+\rangle,\lvert \mathbb{B}_-\rangle , \lvert\downarrow \downarrow\rangle\}$, which can also be written as diag$\{1,1,1,-1\}$ in the basis of $\{\lvert\uparrow \uparrow\rangle, \lvert \uparrow \downarrow\rangle,\lvert \downarrow \uparrow \rangle, \lvert\downarrow \downarrow\rangle\}$, i.e., the canonical CZ gate.

The Hamiltonian~(\ref{H_iswap}) looks like involving only four states $\lvert \pm\rangle$ and $\lvert\mathbb{B}_\pm\rangle$, but Eq.~(\ref{3dAQRM}) shows that it actually couples motional states of different quanta which can result in undesired entanglement between the internal and motional states of the molecules. To show the essence of this coupling, we take only the $z$ mode in Eq.~(\ref{3dAQRM}) because it has the strongest coupling coefficient if in the condition of, e.g., Ref.~\cite{ruttley_long-lived_2025}. Further, we take the initial motional state $\hat{a}_z^\dag \lvert\text{vac}\rangle$ as an example. If we restrict change of motional quantum up to $\pm2$, then the Hamiltonian for the initial state $\lvert\mathbb{B}_+\rangle\otimes\hat{a}_z^\dag \lvert\text{vac}\rangle$ is
\begin{eqnarray}
\hat{H}_+ &=&\left(\begin{array}{ccccc} \eta-\hbar\omega_z&0&\sqrt{2}\zeta_z &g&0\\
0& \eta+2\hbar\omega_z&\sqrt{3}g&\sqrt{6}\zeta_z&0\\\sqrt{2}\zeta_z&
                     \sqrt{3}g&\eta+\hbar\omega_z&\sqrt{2}g&0\\
                     g&  \sqrt{6}\zeta_z&\sqrt{2}g&\eta  &
 \hbar\frac{\Omega_{\mu}(t)}{2}\\
                     0& 0  &0&
 \hbar\frac{\Omega_{\mu}^\ast(t)}{2}&0
                    \end{array}\right)\nonumber\\&&+\text{diag}\{1,7,5,3,0\}\zeta_z,
                    \label{H+case}
\end{eqnarray}
where we have ignored a constant $3\hbar\omega_z/2$ for it induces a trivial overall phase to all the input states, and the last line above is from the term $(2\hat{a}_z^\dag\hat{a}_z+1)\zeta_z$. The first four states in the basis of Eq.~(\ref{H+case}) are $\lvert+\rangle\otimes\{1,\frac{(\hat{a}_z^\dag)^3}{\sqrt{6}} ,\frac{(\hat{a}_z^\dag)^2}{\sqrt{2}} ,\hat{a}_z^\dag \}\lvert\text{vac}\rangle$, and the last is the initial state.

\begin{figure}
\includegraphics[width=3.0in]
{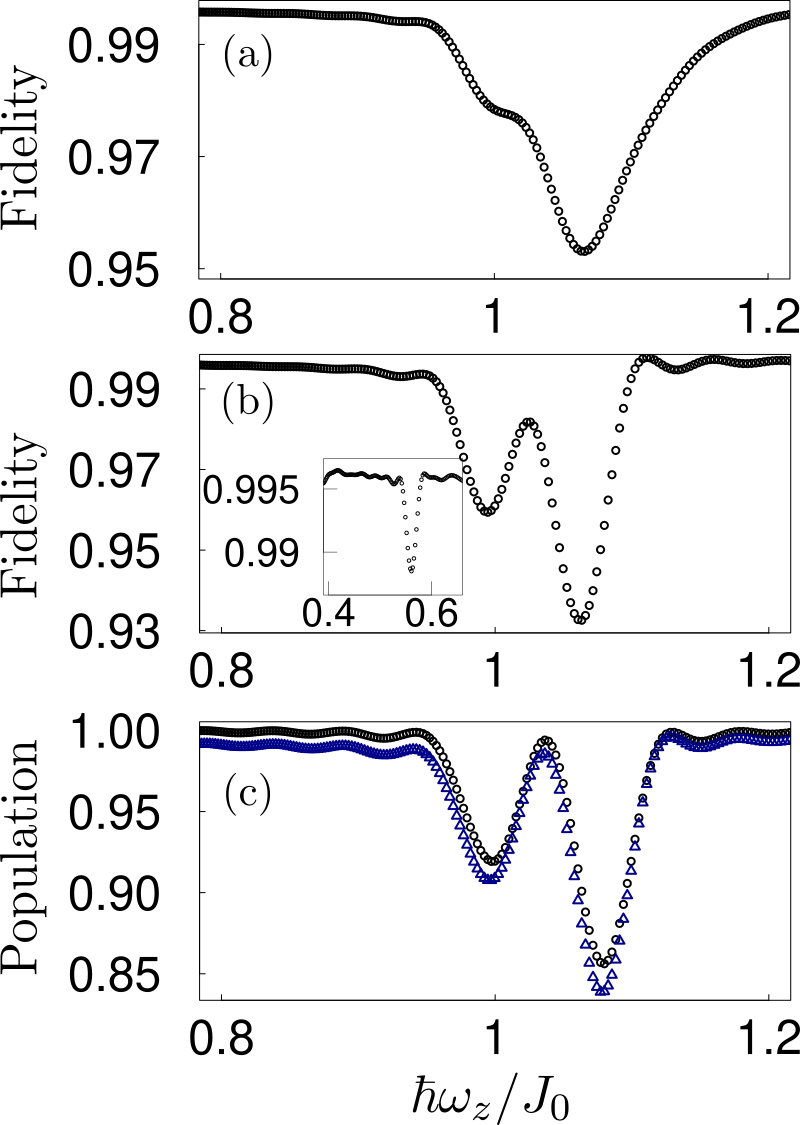}
\caption{(a,b) Trap-dipole resonance signified by the fidelity drop of the CZ gate described around Eq.~(\ref{H_iswap}) when varying $\omega_z$ with $J_0=20\Omega_\mu$ fixed. Here, $(g,\zeta_z)=-(3,6\ell_z/L)\ell_z/L$ and $\ell_z/L=0.045$ as from around Eq.~(\ref{3dAQRM}), and the initial motional state is a thermal state with $\langle \hat{a}_z^\dag\hat{a}_z\rangle=2$ in (a) and $\hat{a}_z^\dag \lvert\text{vac}\rangle$ in (b). The two minima of the gate fidelity in (b) are at $\hbar\omega_z\approx -\eta+\zeta_z$ and $-\eta-5\zeta_z+\frac{3g^2}{\hbar\omega_z+2\zeta_z}$, respectively, where the latter differs from the expected $-\eta-5\zeta_z$ because of the AC stark shift from the off-resonant coupling of strength $\sqrt{3}g$ in Eq.~(\ref{H+case}). The inset shows the resonance at around $J_0=2\hbar\omega_z$. The simulation is via QuTip~\cite{Johansson_2012,Johansson_2013}, and motional states with up to 40 motional excitations of the $\hat{a}_z$ mode are included. Due to the finiteness of the blockade condition~\cite{Shi2021qst}, the gate fidelity is 0.9969 even in the case of no coupling between DDI and molecular motion, i.e., if $\ell_z=0$. (b) The round~(triangle) symbols show the final population in $\hat{a}_z^\dag\lvert\text{vac}\rangle$~($\lvert\uparrow\downarrow\rangle$) after tracing over the motional~(internal) degrees of freedom for the initial state $\lvert\uparrow\downarrow\rangle\otimes\hat{a}_z^\dag\lvert\text{vac}\rangle$. The two minima in (b) and (c) around $J_0=\hbar\omega_z$ do not match well because the fidelity is influenced not only by the population, but also by the phase of the state.  \label{figure-trap-dipole} }
\end{figure}

The trap-dipole resonance is as follows. The desire is that the only allowed transition is from the initial state to $\lvert+\rangle\otimes \hat{a}_z^\dag \lvert\text{vac}\rangle$. However, when $\eta+\hbar\omega_z+5\zeta_z=0$, the states $\lvert+\rangle\otimes\{ \frac{(\hat{a}_z^\dag)^2}{\sqrt{2}} ,\hat{a}_z^\dag \}\lvert\text{vac}\rangle$ and $\lvert\mathbb{B}_+\rangle\otimes \hat{a}_z^\dag \lvert\text{vac}\rangle$ form a resonant ladder-type transition. In the extreme case when $|\eta|\gg \hbar|\Omega_\mu|,|g|$, a resonant transition from the initial state to $\lvert+\rangle\otimes \frac{(\hat{a}_z^\dag)^2}{\sqrt{2}} \lvert\text{vac}\rangle$ emerges with an effective two-photon Rabi frequency $\sqrt{2}g\Omega_\mu/(\eta+3\zeta_z)$~\cite{Shi2014}. Similarly, if $\eta+2\hbar\omega_z+7\zeta_z=0$, a resonant transition can occur from the initial state to $\lvert+\rangle\otimes \frac{(\hat{a}_z^\dag)^3}{\sqrt{6}} \lvert\text{vac}\rangle$ with an effective two-photon Rabi frequency $\sqrt{6}\zeta_z\Omega_\mu/(\eta+3\zeta_z)$. The case for the initial state $\lvert\mathbb{B}_-\rangle $ differs in that the resonance only occurs between the initial state and $\lvert+\rangle\otimes \lvert\text{vac}\rangle$ since $\eta<0$. If the initial motional state is at the ground state, then the resonance exists only for the state $\lvert\mathbb{B}_+\rangle$.

To examine the physical picture shown above, we have numerically simulated the fidelity of the gate shown around Eq.~(\ref{H_iswap}), with results shown in Fig.~\ref{figure-trap-dipole}(a) and (b), with the initial motional state as a thermal state with $\langle \hat{a}_z^\dag\hat{a}_z\rangle=2$ in (a) and $\hat{a}_z^\dag \lvert\text{vac}\rangle$ in (b). The gate fidelity adopts the definition of Ref.~\cite{Pedersen2007} which can capture both phase and population error, where the change of the motional state is accounted for by calculating the inner product of the initial motional state and the actual motional state. Note that in case the initial motional state is pure, the motional wavefunction will acquire a phase factor under the Hamiltonian~(\ref{H-trap01}), though this phase factor does not appear in the case of thermal motional state. Although not obvious in Fig.~\ref{figure-trap-dipole}(a), One can find two trap-dipole resonances around $\hbar\omega_z=J_0$ in Fig.~\ref{figure-trap-dipole}(b). The exact location of the right resonance is shifted from the expected location by about $3g^2/(\hbar\omega_z)$, which is due to that in Eq.~(\ref{H+case}), there is an off-resonant coupling between the two states $\lvert+\rangle\otimes \frac{(\hat{a}_z^\dag)^3}{\sqrt{6}} ,\frac{(\hat{a}_z^\dag)^2}{\sqrt{2}}   \}\lvert\text{vac}\rangle$, which shifts the energy of the latter by $3g^2/(\hbar\omega_z)$. Due to the much smaller coupling $\zeta_z$ that changes the motional quantum by 2, one can find in Fig.~\ref{figure-trap-dipole}(a) that the resonant behaviour at around $\hbar\omega_z\approx J_0/2$ is much weaker.

To examine whether the fidelity loss is due to coupling to unwanted motional states, the round symbols in Fig.~\ref{figure-trap-dipole}(c) show the
final population of the motional state in $\hat{a}_z^\dag\lvert\text{vac}\rangle$ for the initial state $\lvert\uparrow\downarrow\rangle\otimes\hat{a}_z^\dag\lvert\text{vac}\rangle$. One can see that the drop of the population in the initial motional state has a similar tendency of the fidelity drop as in Fig.~\ref{figure-trap-dipole}(b). Notably, at $\hbar\omega_z/J_0\approx1.08$, the populations in states of zero, one, two, and three motional quanta of the $\hat{a}_z$ mode are $0.002, 0.856,~0.134$, and $0.007$, respectively, indicating that the loss is mainly to the state with two motional quanta. Because the states with altered motional quantum states is not resonant during the second microwave pulse for bringing the population back to the ground state, there will be corresponding population error in the computational basis state, shown by the triangle symbols of Fig.~\ref{figure-trap-dipole}(c).

We have analyzed the trap-dipole resonance considering the motion along the molecular separation axis. This trap-dipole resonance will also occur for the axial motion when $\hbar\omega_x\approx J_0/2$. Due to the similar mechanism, we do not show results for the trap-dipole resonance for the axial mode.

\section{Modified iSWAP}~\label{sec-swap}
The simplest way to use DDI of polar molecules for quantum gates is to use the exchange process, which can realize an iSWAP gate~\cite{Ni_2018}. The standard way to realize this gate is to use two $\pi$ pulses of microwave field to toggle the interaction~\cite{picard_entanglement_2025} in the $\pi$-wait-$\pi$ sequence, which can be nearly perfect if $J/(\hbar\Omega_\mu)\rightarrow0$. Due to that it is better not to have too large $\Omega_\mu$ in order to avoid exciting off-resonant transitions~\cite{Ni_2018}, it seems the theoretical fidelity is limited. Below, we show that with finite $J/(\hbar\Omega_\mu)$ or even with $J/(\hbar\Omega_\mu)>1/2$, a high-fidelity iSWAP can be realized. Moreover, one microwave pulse can realize the iSWAP.

\subsection{The iSWAP gate with control sequence of Ref.~\cite{picard_entanglement_2025}}
It was theoretically proposed in Ref.~\cite{Ni_2018}, and experimentally demonstrated in Ref.~\cite{picard_entanglement_2025} that an iSWAP gate can be realized with a pair of molecules. The molecules are microwave excited to states so that the flip-flop interaction takes place, causing an input-dependent exchange of the internal states of the two molecules. Briefly, this type of iSWAP gate works as follows. For the four input states
$\{\lvert\uparrow \uparrow\rangle,  \lvert\mathbb{B}_+\rangle,\lvert \mathbb{B}_-\rangle , \lvert\downarrow \downarrow\rangle\}$, a $\pi$ pulse of the microwave field via the Hamiltonian~(\ref{H_iswap}) with $J/(\hbar\Omega_\mu)\ll1$ will change them to $\{\lvert\uparrow \uparrow\rangle,  -i\lvert+\rangle, -i\lvert  -\rangle , -\lvert e, e\rangle\}$, where $\lvert\pm\rangle$ are defined below Eq.~(\ref{Rabi-02}). After a wait duration $\pi\hbar/(2|J|)$, the two states $ -i\lvert+\rangle$ and $-i\lvert  -\rangle  $ become $\lvert+\rangle$ and $-\lvert  -\rangle$, respectively because of the attractive DDI. A second microwave $\pi$ pulse then restores $\lvert+\rangle,~-\lvert  -\rangle$ and $-\lvert e, e\rangle$ back to ground states $-i\lvert\mathbb{B}_+\rangle,i\lvert \mathbb{B}_-\rangle$ and $\lvert\downarrow \downarrow\rangle$. As a whole, this $\pi$-wait-$\pi$ sequence results in a map diag$\{1,-i,i,1\}$ in the basis $\{\lvert\uparrow \uparrow\rangle,  \lvert\mathbb{B}_+\rangle,\lvert \mathbb{B}_-\rangle , \lvert\downarrow \downarrow\rangle\}$, which is equivalent to
\begin{eqnarray}
\lvert\uparrow \uparrow\rangle&\rightarrow& \lvert\uparrow \uparrow\rangle,\nonumber\\
\lvert\uparrow \downarrow\rangle&\rightarrow& -i\lvert\downarrow \uparrow\rangle,\nonumber\\
\lvert\downarrow \uparrow\rangle&\rightarrow& -i\lvert\downarrow \uparrow\rangle,\nonumber\\
\lvert\downarrow \downarrow\rangle&\rightarrow& \lvert\downarrow \downarrow\rangle
                    \label{i-gate}
\end{eqnarray}

\begin{figure}
\includegraphics[width=3.0in]
{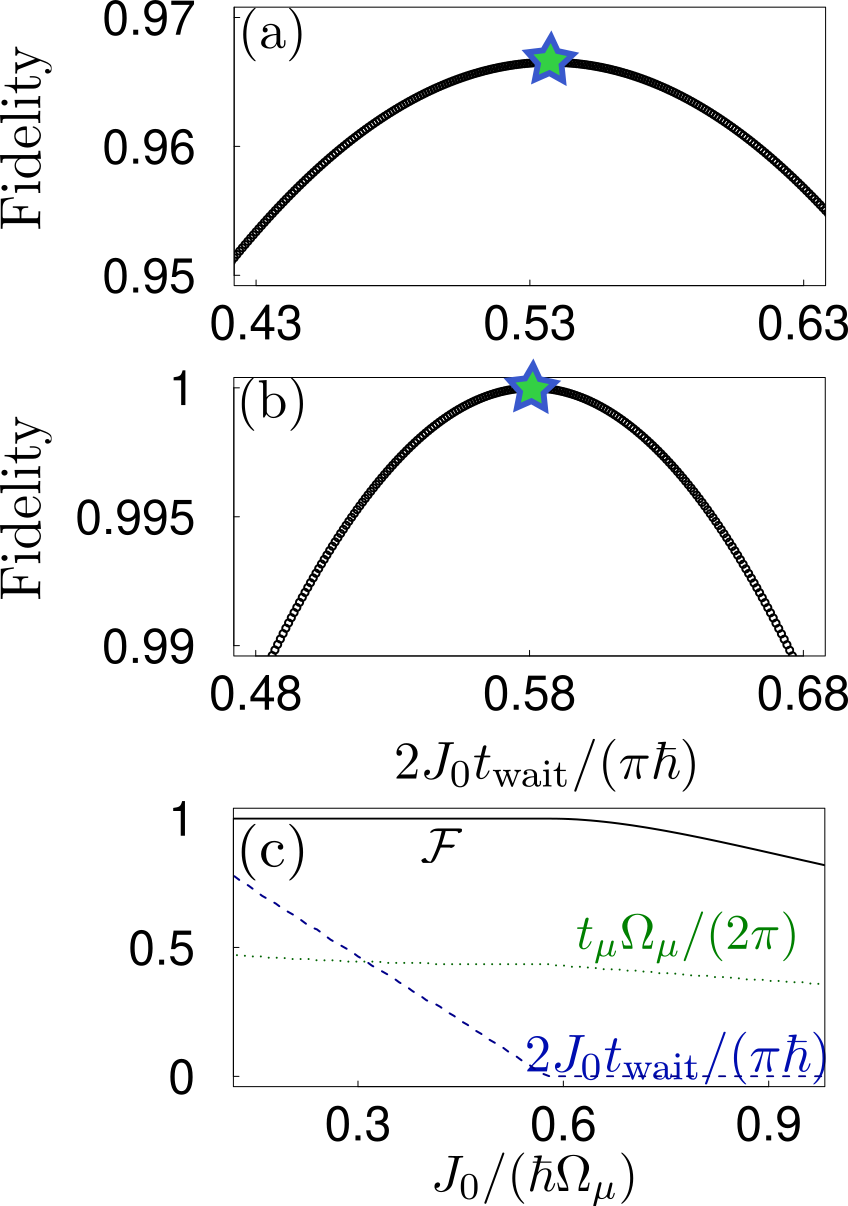}
\caption{(a) Fidelity of the standard $\pi$-wait-$\pi$ iSWAP gate~\cite{picard_entanglement_2025} as a function of the wait time when the pulse area of the microwave pulses is $\pi$. The maximal fidelity is $0.9665$ at the star. (b) When the pulse area of the microwave pulses is $0.906\pi$, the fidelity of the modified iSWAP can reach 1, marked by the star. In (a,b), $\Omega_\mu=2\pi\times1.6$~kHz and $J_0=h\times0.37$~kHz. (c) Fidelity, optimal pulse duration $t_\mu$ and wait duration as a function of $J_0/(\hbar\Omega_\mu)$ when $J_0$ is fixed at $h\times 0.37$~kHz. The optimal fidelity is unit when $J_0/(\hbar\Omega_\mu)\in(0.05,~0.58)$, with $t_{\text{wait}}$ decreases to 0 at $J_0/(\hbar\Omega_\mu)=0.573$.
\label{figure-swap-op} }
\end{figure}

The gate in Eq.~(\ref{i-gate}) depends on the condition of $J/(\hbar\Omega_\mu)\rightarrow0$, but in practice it is finite. Then, the DDI already occurs during the two $\pi$ pulses, which makes it necessary to shorten the wait time $t_{\text{wait}}$ from the expected value. In the iSWAP gate with two NaCs molecules of Ref.~\cite{picard_entanglement_2025}, $\Omega_\mu$ is about $2\pi\times1.6$~kHz, the molecular-frame dipole moment is \textdong~$=4.6$~debye~\cite{Aymar_2005}, and the measured trap separation is $L=1.9~\mu$m. Using the same quantum chemistry calculation, the theoretical value of \textdong~reported in Ref.~\cite{Aymar_2005} was reported again in Ref.~\cite{Deiglmayr_2008}. This should yield $J_0=h\times0.31$~Hz. However, in the analyses of their data, Ref.~\cite{picard_entanglement_2025} used $L=1.79~\mu$m in order to match the observed interaction rate. Therefore, we would also take this latter $L$ to analyze the gate, at which $J_0$ is about $h\times 0.37$~kHz. This set of parameters does not satisfy $J_0/(\hbar\Omega_\mu)\ll1$, but one can numerically find an optimal wait duration $t_{\text{wait}}=0.537\frac{\pi\hbar}{2J_0}\approx360~\mu$s when the pulse duration $t_\mu$ for both microwave pulse is $\pi/(2\Omega_\mu)$, at which the gate fidelity is maximal, equal to $0.9665$. The gate fidelity as a function of $2J_0t_{\text{wait}}/(\pi\hbar)$ is shown in Fig.~\ref{figure-swap-op}(a). The value of 360$~\mu$s found here deviates from the wait duration 330~$\mu$s employed in Ref.~\cite{picard_entanglement_2025}, which is possibly due to that the actual value of $L$ is not equal to $1.79~\mu$m, or due to that the actual value of ~\textdong~ is yet to be confirmed for its value of Ref.~\cite{picard_entanglement_2025} quoted from the theory in Ref.~\cite{Aymar_2005} differs by 3\% from, e.g., the experimental result of Ref.~\cite{Dagdigian_1972}.

\subsection{High-fidelity iSWAP gate with modified control sequence}
We find that if the two microwave pulses are not $\pi$ pulses, but a shorter pulse, then the gate fidelity can even be larger. Numerically, one needs to locate an optimal combination of the pulse duration $t_\mu$ and the wait duration $t_{\text{wait}}$: when $t_\mu=0.906\frac{\pi}{\Omega_\mu}$ and $t_{\text{wait}}=0.581\frac{\pi\hbar}{2J_0}$, the gate fidelity $\mathcal{F} $ is 1 with $\Omega_\mu=2\pi\times1.6$~kHz and $J_0=h\times 0.37$~kHz. The fidelity as a function of the wait duration is shown in Fig.~\ref{figure-swap-op}(b) when $t_\mu=0.906\frac{\pi}{\Omega_\mu}$. This means that due to the modification of the DDI in the Rabi oscillation between the ground and the excited state, a higher fidelity is achievable with the microwave pulse area smaller than $\pi$.

Further, if we vary the magnitude of DDI or $\Omega_\mu$, perfect gate fidelity can also be achieved. It was shown that too large $\Omega_\mu$ or too strong DDI will cause off-resonant excitation to nearby states, resulting in extra control error~\cite{Ni_2018}. Therefore, we fix the DDI, but vary $\Omega_\mu$, and find that there are indeed conditions to have perfect gates with $J_0/(\hbar\Omega_\mu)$ up to about 0.6, shown in Fig.~\ref{figure-swap-op}(c). Importantly, there are even conditions to realize a perfect gate with $t_{\text{wait}}=0$. For example, Fig.~\ref{figure-swap-op}(c) shows that with a total pulse duration $2t_\mu=1.732\frac{\pi}{\Omega_\mu}$ with $\Omega=2\pi\times0.641$~kHz, the gate fidelity is 1. Since the gate without any wait duration is simple to implement in experiment, we analyze it in more detail below.

\subsection{Influence of QM-DDI coupling on the one-pulse iSWAP gate}
\begin{figure}
\includegraphics[width=3.0in]
{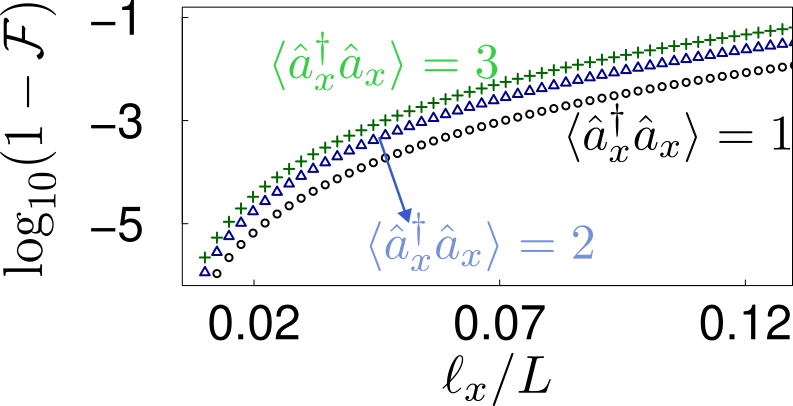}
\caption{Logarithm of the infidelity of the one-pulse iSWAP gate as a function of $\ell_x/L$ with $\Omega_\mu = 2\pi\times0.641$~kHz. Here, $J_0=h\times 0.37$~kHz as analyzed below Eq.~(\ref{i-gate}) according to the setup of Ref.~\cite{picard_entanglement_2025}. The round, triangle, and cross symbols are data when the thermal motional state has one, two, and three quanta in the $\hat{a}_x$ mode. To avoid trap-dipole resonance, we set $\omega_x= 2\Omega_\mu$ so that the infidelity here is only due to the bare QM-DDI coupling. As in Fig.~\ref{figure-trap-dipole}, motional states with up to 40 motional excitations of the $\hat{a}_x$ mode are included in the quantum simulation of the in-trap motion via QuTip~\cite{Johansson_2012,Johansson_2013}. Here, we show results below $\ell/L=0.12$ because it is the largest value from recent experiments~\cite{ruttley_long-lived_2025}. For the experiment in Ref.~\cite{picard_entanglement_2025} where $\ell_x/L$ is about $0.05$, the gate fidelities are $0.9998,~0.9994$, and $0.9988$ when $\langle\hat{a}_x^\dag\hat{a}_x\rangle=1,~2$, and $3$, respectively.  \label{figure-swap} }
\end{figure}

It is useful to examine how the one-pulse iSWAP gate can tolerate with a nonzero $\ell_\xi$ because the QM-DDI coupling will change the picture of the time dynamics. In particular, during the wait time, the molecular DDI will couple the motional state and the internal states, creating entanglement between the two degrees of freedom because the motional states with different vibration quanta have different coupling strengths. When excited back to the computational basis, the change of the motional state for the input components $\lvert\uparrow \downarrow\rangle,
\lvert\downarrow \uparrow\rangle$ compared to the input components
$\lvert\uparrow \uparrow\rangle, \lvert\downarrow \downarrow\rangle$ induces a control error. To examine this fidelity loss, we take the trap geometry in Ref.~\cite{picard_entanglement_2025} as an example because an iSWAP gate was demonstrated therein. Two nearby molecules were trapped in tweezers with a strong axial trap frequency $\omega_x/(2\pi)=4.95$~kHz in Ref.~\cite{picard_entanglement_2025}, which is over 12 times stronger than that of Ref.~\cite{ruttley_long-lived_2025}. It was shown in Ref.~\cite{picard_entanglement_2025} that the only relevant motion of the molecule is along the axial direction, with $\ell_x\approx0.11~\mu$m and $L\in[1.79,~2.5]~\mu$m, yielding $\ell_x/L\in(0.04,~0.06)$. Therefore, we assume that the motion along the radial direction is negligible, and numerically studied the gate infidelity, with the initial motional state in a thermal equilibrium possessing one, two, and three motional quanta in the $\hat{a}_x$ mode in Fig.~\ref{figure-swap}, respectively. It shows that with tightly confined molecules as in Ref.~\cite{ruttley_long-lived_2025} where $\langle \hat{a}_x^\dag\hat{a}_x\rangle\approx2$ and $\ell_x/L\approx0.05$, the gate fidelity is about 0.9994. In principle, the inhomogeneous phase accumulation in different motional states can cause a control error, which is present if the initial motinal state is not thermal. This can cause a large error especially the phase of the output state should be the desired one in a quantum logic gate. However, numerical results with pure motional states show that the fidelity drop is similar as in Fig.~\ref{figure-swap}. This indicates that the population in excited motional states, instead of the coherence of the motional states, is the main cause of the control error.

\section{Two-qubit controlled-phase gates}~\label{sec-quasi}
The blockade mechanism as studied in Sec.~\ref{sec-resoance} is of specific applicability, but the blockade-mediated gate has an intrinsic error $\sim (\hbar\Omega_\mu/J)^2$ due to the finiteness of the DDI~\cite{Shi2017}, so that a sufficient large $J/(\hbar\Omega_\mu)$ is needed to reach a high fidelity. However, due to the difficulty to place two molecules too close, a high-fidelity blockade molecular gate may need a Hz-scale $\Omega_\mu$, but meanwhile the decoherence rate of the qubit state is also on the Hz scale~\cite{Cornish_2024}.

Therefore we would introduce an alternative blockade-like gate, and consider the following microwave driving
\begin{eqnarray}
\hat{H}_{\mu}(\Omega) = \left[\hbar\frac{\Omega}{2}\lvert e\rangle\langle\uparrow\rvert +\text{H.c.} \right],\label{H_mu-2}
\end{eqnarray}
which differs from Eq.~(\ref{H_mu-1}) in that the $\uparrow$ state is excited, except of the $\downarrow$ state as in Eq.~(\ref{H_mu-1}). For brevity, we assume $\Omega$ to be real and positive.

Below, we describe a two-qubit controlled phase gate with an arbitrary phase $\varphi$. The gate is like a blockade gate~\cite{Shi2021qst}, but only with $J/(\hbar\Omega_\mu)$ around 10 to have a perfect gate fidelity if no QM-DDI occurs. It needs eight pulses, each with a pulse duration $t_{\mu}=\pi/(2\Omega)$, i.e., with a pulse area $\pi/2$, so that the total required pulse area is $4\pi$. For brevity, we assume that the pulse is square. Before showing the gate, it is useful to show a simple two-pulse gate to explain the essence.

\subsubsection{A gate with a $\pi-\pi$ pulse sequence}\label{pi-pi-gate}
Because the state $\lvert \downarrow\downarrow\rangle$ is not excited, it stays intact in the rotating frame. Below, we describe the state dynamics for $\lvert \uparrow\downarrow\rangle$, $\lvert \downarrow\uparrow \rangle$, and $\lvert \uparrow\uparrow\rangle$.

The time dynamics for either the input state $\lvert \uparrow\downarrow\rangle$ or $\lvert \downarrow\uparrow\rangle$ is simply a resonant Rabi oscillation. For the input state $\lvert \uparrow\uparrow\rangle$, the Hamiltonian is
\begin{eqnarray}
\hat{H}_{\uparrow\uparrow}(\Omega) &=&\left(\begin{array}{ccc}
                0           & \hbar\frac{\Omega}{\sqrt{2}}&0\\
                 \hbar\frac{\Omega^\ast}{\sqrt{2}}  & J &
 \hbar\frac{\Omega}{\sqrt{2}} \\
                  0   &
 \hbar\frac{\Omega^\ast}{\sqrt{2}}&0
                    \end{array}\right)
                    \label{Hpm-gate1}
\end{eqnarray}
in the basis of
\begin{eqnarray}
 \{\lvert ee\rangle  ~,  \lvert+\rangle   ~,\lvert \uparrow\uparrow\rangle\},\nonumber
\end{eqnarray}
where $\lvert+\rangle\equiv (\lvert\uparrow~e\rangle+ \lvert~e\uparrow\rangle)/\sqrt{2}$. In the limit of $|J|\gg \Omega$, one can adiabatically eliminate the intermediate state $\lvert+\rangle$, reaching an effective Hamiltonian $(\Omega_{\text{eff}}\lvert ee\rangle\langle \uparrow\uparrow\rvert/2+ $H.c.)$+\hat{H}_{\text{AC}}$, where $\Omega_{\text{eff}}= -\Omega^2/(2J)$ and the AC stark shift $\hat{H}_{\text{AC}}$ has a term $-|\Omega|^2/(4J)\lvert\uparrow\uparrow\rangle\langle \uparrow\uparrow\rvert $~\cite{Shi2014}. If $|J|$ is sufficiently large, the AC stark shift can be neglected. Then, a $\pi$ pulse with Rabi frequency $\Omega$ can excite
$\lvert \uparrow\downarrow\rangle$ and $\lvert \downarrow\uparrow\rangle$ to $-i\lvert e\downarrow\rangle$ and $-i\lvert \downarrow e\rangle$, respectively. Then, a second $\pi$ pulse with Rabi frequency $i\Omega$ can excite $-i\lvert e\downarrow\rangle$ and $-i\lvert \downarrow e\rangle$ back to $i\lvert \uparrow\downarrow\rangle$ and $i\lvert \downarrow\uparrow\rangle$, respectively. Meanwhile, because of the $\pi/2$ phase change in the Rabi frequency, the effective two-photon Rabi frequency for the input state $\lvert \uparrow\uparrow\rangle$ becomes $\Omega_{\text{eff}}= -\Omega^2/(2J)$, i.e., exactly opposite to that in the first pulse. As a result, the input state $\lvert \uparrow\uparrow\rangle$ evolves back to the original state as in the spin-echo sequence~\cite{Shi2018prapp2}, leading to a gate of map diag$\{ \lvert \uparrow\uparrow\rangle, ~i\lvert \uparrow\downarrow\rangle, ~i\lvert \downarrow\uparrow\rangle, ~\lvert \downarrow\downarrow\rangle\}$. This gate is equivalent to the CZ gate by a phase change $\pi$ to the $\downarrow$ state of both qubits. Due to the presence of the AC stark shift, however, there is an extra phase shift to $\lvert \uparrow\uparrow\rangle$, resulting in a gate fidelity $0.9903$ if $J/(\hbar\Omega_\mu)=10$. We find that the infidelity is mainly due to the AC stark shift because when we replace $J/(\hbar\Omega_\mu)=10$ by $J/(\hbar\Omega_\mu)=9.798$, the spin-echo population restoration is perfect~\footnote{As shown in Ref.~\cite{Shi2018prapp2}, the population leakage in a detuned Rabi oscillation is sensitive to the detuning.}, but the gate fidelity is still small, only equal to $0.9900$.
\begin{figure}
\includegraphics[width=2.50in]
{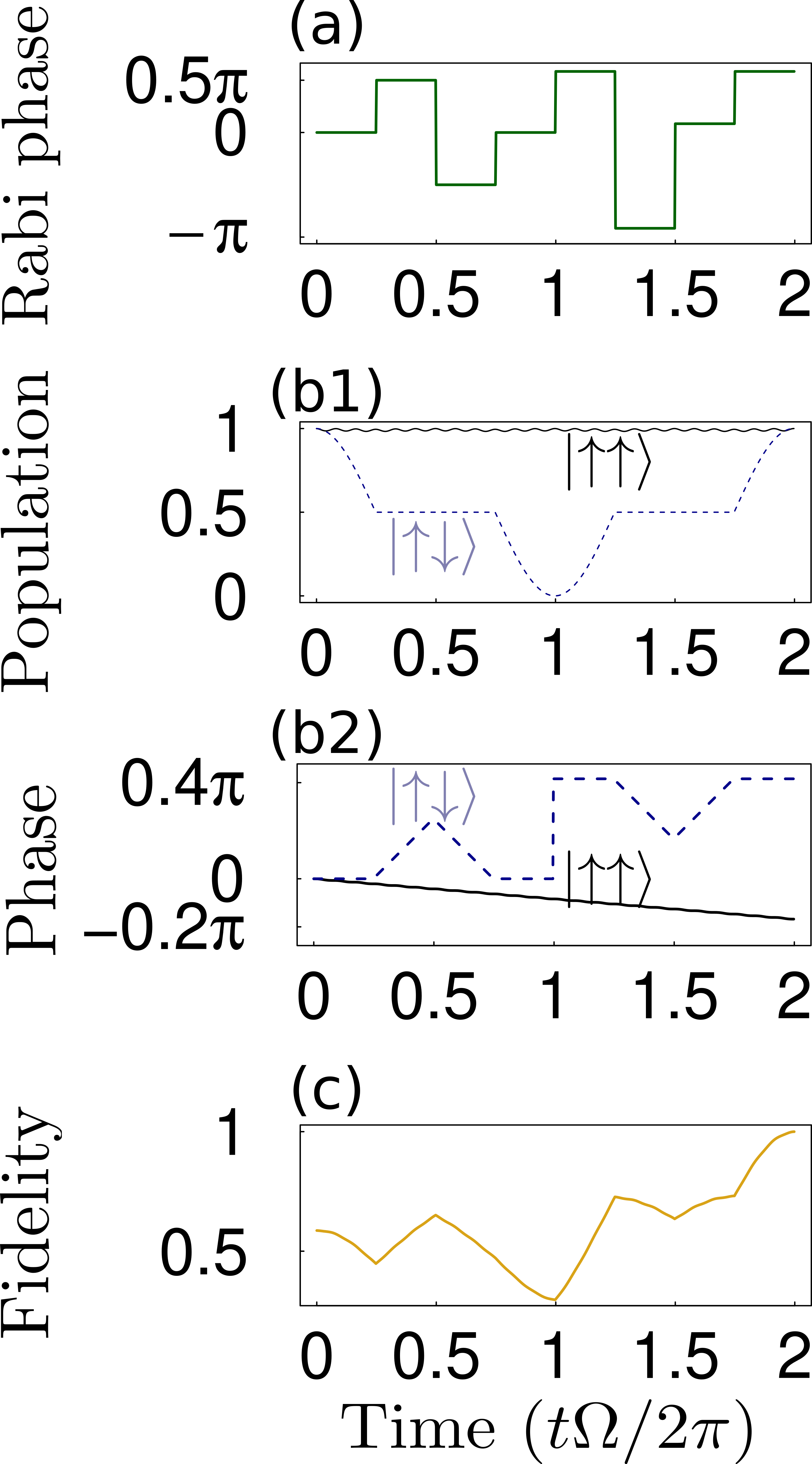}
\caption{(a) Phase of the microwave Rabi frequency for implementing the quasi-blockade gate with $J/(\hbar\Omega)=11.832$. In (b1,b2), the population and phase of the input state component are shown by the solid~(dashed) curves if the input state is $\lvert\uparrow\uparrow\rangle~(\lvert\uparrow\downarrow\rangle)$, respectively. The final phases are $-0.16785\pi$ and $0.41608\pi$ in $\lvert\uparrow\uparrow\rangle$ and $\lvert\uparrow\downarrow\rangle$, respectively. The state dynamics for $\lvert\downarrow\uparrow\rangle$ is similar to that of $\lvert\uparrow\downarrow\rangle$. For the input state $\lvert\uparrow\uparrow\rangle$, its population nearly stays constant, which indicates that the DDI-coupled state $\lvert+\rangle$ is barely excited. (c) Time evolution of the fidelity of the quasi-blockade gate.   \label{figure-quasi} }
\end{figure}
\subsubsection{A high-fidelity quasi-blockade gate}
To cope with the AC stark shift, we extend the above $\pi-\pi$ gate to an eight-pulse gate, each pulse with an pulse area of $\pi/2$. The magnitudes of the Rabi frequencies are all equal to $\Omega$, but with a phase of $\{0,\pi/2,-\pi/2,0\}$ for the first four pulses, and $\{0,\pi/2,-\pi/2,0\}-2\vartheta+\pi/2$ for the fifth to the eighth pulses, where $\vartheta$ is equal to the theoretical phase shift for the input state $\lvert\uparrow\uparrow\rangle$ during each pair of two consecutive pulses. According to the physical picture shown in Sec.~\ref{pi-pi-gate}, the state evolution for the input state $\lvert\uparrow\uparrow\rangle$ is always a spin-echo sequence for each two consecutive pulses, resulting in a phase shift $\vartheta$ only. In other words, after the eight pulses, we have \begin{eqnarray}
 \lvert \uparrow\uparrow\rangle\mapsto e^{4i\vartheta}\lvert \uparrow\uparrow\rangle.
\end{eqnarray}
 For the input state $\lvert\uparrow\downarrow\rangle$ or $\lvert\downarrow\uparrow\rangle$, only the $\uparrow$ component evolves, which is as follows,
\begin{eqnarray}
 \lvert \uparrow\rangle &\xrightarrow[]{\Omega}&  \frac{\lvert \uparrow\rangle-i\lvert e\rangle}{\sqrt{2}}\xrightarrow[]{i\Omega}\frac{e^{i\frac{\pi}{4}}\lvert \uparrow\rangle+e^{-i\frac{\pi}{4}}\lvert e\rangle}{\sqrt{2}}\xrightarrow[]{-i\Omega} \frac{\lvert \uparrow\rangle-i\lvert e\rangle}{\sqrt{2}} \nonumber\\&\xrightarrow[]{\Omega}&
 -i\lvert e\rangle   \xrightarrow[]{i\Omega e^{-2i\vartheta}} i\frac{ e^{2i\vartheta}\lvert \uparrow\rangle-\lvert e\rangle}{\sqrt{2}}
   \xrightarrow[]{-\Omega e^{-2i\vartheta}} \nonumber\\ && e^{ i\frac{\pi}{4}}\frac{ e^{2i\vartheta}\lvert \uparrow\rangle-\lvert e\rangle}{\sqrt{2}} \xrightarrow[]{\Omega e^{-2i\vartheta}}i\frac{ e^{2i\vartheta}\lvert \uparrow\rangle-\lvert e\rangle}{\sqrt{2}}\nonumber\\ && \xrightarrow[]{i\Omega e^{-2i\vartheta}} i e^{2i\vartheta}\lvert \uparrow\rangle,
\end{eqnarray}
where each arrow denotes one pulse. As a result, we have the following state map with the eight pulses,
 \begin{eqnarray}
 \lvert \uparrow\uparrow\rangle&\mapsto& e^{4i\vartheta}\lvert \uparrow\uparrow\rangle,\nonumber\\
 \lvert \uparrow\downarrow\rangle& \mapsto&  i e^{2i\vartheta}\lvert \uparrow\downarrow\rangle,\nonumber\\
 \lvert \downarrow \uparrow\rangle& \mapsto&  i e^{2i\vartheta}\lvert \uparrow \uparrow\rangle,\nonumber\\
 \lvert \downarrow\downarrow\rangle& \mapsto & \lvert \downarrow\downarrow\rangle,\label{bloacke-gate01}
\end{eqnarray}
which is equivalent to the CZ gate with a single-qubit phase gate $\lvert \uparrow\rangle \mapsto  -i e^{-2i\vartheta}\lvert \uparrow\rangle$ for both qubits.

\begin{figure}
\includegraphics[width=3.0in]
{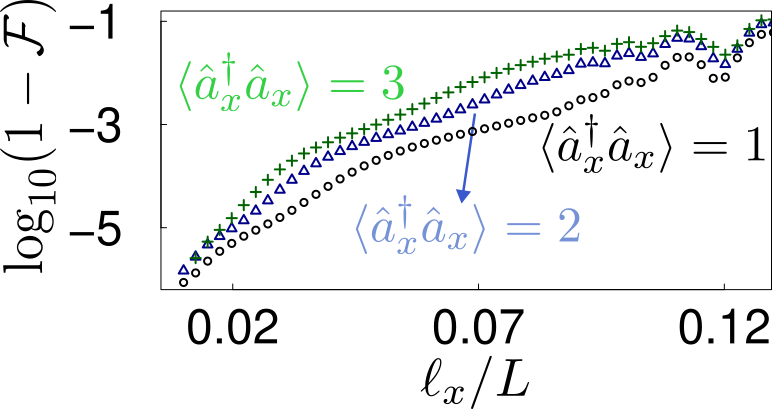}
\caption{Logarithm of the infidelity of the quasi-blockade gate as a function of $\ell_x/L$ with the parameters of Fig.~\ref{figure-quasi} except of an extra parameter $\omega_x= 0.3\Omega_\mu$ here. The round, triangle, and cross symbols are data when the thermal motional state has one, two, and three quanta in the $\hat{a}_x$ mode. When $\ell_x/L=0.05$, the gate fidelities are $0.9998,~0.9995$, and $0.9990$ when $\langle\hat{a}_x^\dag\hat{a}_x\rangle=1,~2$, and $3$, respectively. \label{figure-blockade} }
\end{figure}

One can find that if the phase of the Rabi frequencies for the fifth to the eighth pulses are $\{0,\pi/2,-\pi/2,0\}-2\vartheta+\varphi/2$, then the two factors $i$ in the second and third lines in Eq.~(\ref{bloacke-gate01}) are updated to $-e^{-i\varphi/2}$. Then, the gate becomes an arbitrary two-qubit controlled phase gate $C(\varphi)$ with map diag$\{e^{i\varphi},~1,~1,~1\}$ up to single-qubit phase change to both qubits. The phase $\varphi$ can be tuned to any desired value simply by adjusting the phase of the microwave fields.

Because we are interested in the regime of a moderately large $J/(\hbar\Omega_\mu)$, the adiabatic condition to eliminate the intermediate state discussed around Eq.~(\ref{Hpm-gate1}) is not perfect. However, we find with $J/(\hbar\Omega_\mu)=11.382$, a perfect spin-echo sequence can be realized with each pair of two pulses shown above, with $\vartheta/\pi =- 0.04196$, resulting in a gate with perfect accuracy as shown in Fig.~\ref{figure-quasi}. The time for the input state $\lvert \uparrow\uparrow\rangle$ to stay in $\lvert +\rangle$ is $0.0139\frac{2\pi}{\Omega}$, which means that there is little population in the DDI-coupled state. This can also be signified by the solid curve in Fig.~\ref{figure-quasi}(b1), which shows that the input state $\lvert \uparrow\uparrow\rangle$ nearly always stays there, barely excited. When including the QM in the DDI, a quantum mechanical simulation of the gate fidelity is shown in Fig.~\ref{figure-blockade} when $\varphi=\pi$, i.e, for the case of a CZ gate. Remarkably, the decrease of the fidelity has a slight oscillation. This is because the quasi-blockade gate mainly depends on the phase accumulation of the input state, which further depends on the value of DDI. The increase of the $\ell_x$ causes an effective decrease of DDI due to the term $-3[(\hat{a}_x^\dag+\hat{a}_x)\ell_x/L]^2$ in Eq.~(\ref{J-fluctuation3}). In the iSWAP gate as studied in Fig.~\ref{figure-swap}, the fidelity drop is monotonic because the population exchange in the input state of the iSWAP gate is more important, which has an error monotonic with small $\ell_x/L$. The fidelity in Fig.~\ref{figure-blockade} is similar to that of the iSWAP gate in Fig.~\ref{figure-swap}, but it doesn't mean that $C(\varphi)$ has no advantage since two iSWAP gates are needed to construct a gate equivalent to CZ~\cite{Williams2011}.

\section{Discussions}\label{sec-06}
\subsection{AQRM}
There is one special feature for the AQRM presented in Sec.~\ref{sec-rabi}, i.e., the bosoic mode is the real space motional mode. This mode is not coupled to free-space degrees of freedom as in cavity QED system, where the bosonic mode is cavity photon which can leak into free space.

With setups as in Refs.~\cite{ruttley_long-lived_2025,picard_entanglement_2025}, the AQRM with the $\sigma_3$ term equal to zero can be simulated~\footnote{To avoid confusion, we use $\sigma_{1,2,3}$ to denote the three Pauli matrices for the subscripts $x,y,z$ are used to denote the three coordinates. } We also show that with appropriate state choice in the molecules, the widely studied AQRM with the $\sigma_3$ term can also be simulated. Moreover, the parameters in the AQRM in Sec.~\ref{sec-rabi} are highly tunable. As shown in Sec.~\ref{sec-rabi-B}, $\Delta$ can be tuned from tens of Hz to several kHz by choosing appropriate hyperfine-Zeeman substates. With the setup of Fig.~\ref{figure-1d}(a), the parameter $g=3\ell_z g/L$ is small because $\ell_z$ is small. But if the axes of the two traps are on one line, then $\ell_z$ will be the largest among the three $\ell_\xi$. Then, we can achieve AQRM with large $g$; of course, it is technically demanding to prepare such tweezers. Finally, $\omega$ can be easily tuned by varying the power of the tweezer light. These mean that polar molecules can realize AQRM with a rich variety of parameter regimes.

\subsection{Trap-dipole resonance}
The axial trap frequency of the tweezer in Ref.~\cite{ruttley_long-lived_2025} was 0.4~kHz, while the magnitude of DDI can be around 1~kHz~\cite{picard_entanglement_2025}, therefore the trap-dipole resonance is readily observable with current setups.

Trap-dipole resonance can be relevant for large-scale array in near future. Because the trap frequency is determined by the power of the laser at the trap, it will be challenging to have a large trap frequency in a large tweezer trap. For example, if the same laser power for creating one tweezer is used for creating 100 tweezers, then the trap frequency will drop by a factor of 10. The axial trap frequencices in Refs.~\cite{ruttley_long-lived_2025,picard_entanglement_2025} were about 400~Hz and 4950~Hz, respectively, while the values of $J/h$ were 7~Hz and 370~Hz, respectively, thus the trap-dipole resonance was well avoided therein. However, one can see that if one would pursue a quantum computer, fast quantum gates are needed, which depends on larger $J$. In this case, a kHz-scale $J$ may be needed, which can be near the axial trap frequency especially if a large-scale tweezer array is to be employed.    

\subsection{Quantum gates}
The modified iSWAP shown in Fig.~\ref{figure-swap} only needs one microwave pulse, with total pulse area of about $1.7\pi$, therefore offers a simple way to use polar molecules for entanglement generation.

In quantum information processing, a two-qubit controlled gate, such as the two-qubit controlled phase gate with a phase $\varphi$, $C(\varphi)$, is of specific applicability. For example, by using the CZ gate, i.e., $C(\pi)$, Ref.~\cite{Graham2022} demonstrated quantum phase estimation for a chemistry problem and the quantum approximate optimization algorithm for the maximum cut graph problem, and Ref.~\cite{Evered_2025} simulated a quantum system with non-Abelian spin-liquid phase. For the iSWAP gate as in Ref.~\cite{picard_entanglement_2025}, two iSWAP gates are needed to construct a CNOT gate which is equivalent to the CZ~\cite{Williams2011}. Moreover, if one can realize a gate with $\varphi$ tunable to be any desired value, then it is of particular strength: such a class of tunable controlled phase gates can be used in quantum Fourier transform for notable quantum algorithms like, e.g., phase estimation, order-finding, and the Shor's algorithm~\cite{Nielsen2000}. This makes ultracold polar molecules an alternative candidate for realizing a universal quantum computer.

It is useful to give a final comment on the gate fidelity. By assuming thermal motional state with two motional quanta in the $\hat{a}_x$ mode as from Ref.~\cite{picard_entanglement_2025}, Figs.~\ref{figure-swap} and~\ref{figure-blockade} show that the gate fidelity is around 0.9994 for the one-pulse iSWAP and the quasi-blockade gate. Note that in principle, the molecules can be cooled to the ground QM state, at which the gate infidelity should be much smaller than those shown in Figs.~\ref{figure-swap} and~\ref{figure-blockade}. These indicate a feasibility to realize a high-fidelity iSWAP or $C(\varphi)$ in near future.

\section{Conclusions}\label{sec-07}
We present a quantum mechanical treatment of the coupling between the in-trap quantized motion~(QM) and the dipole-dipole interaction~(DDI) between two optically trapped polar molecules. There are two consequences from this coupling. (1) The asymmetric quatum Rabi model~(AQRM) has been extensively explored in condensed matter physics and cavity quantum electrodynamics, and also is of great interest in the study of fundamental physics, including quantum integrability. We show that the molecular QM-DDI coupling leads to a novel AQRM where the bosonic mode is the QM of the molecules, bringing new perspectives for the study of AQRM. (2) The QM-DDI coupling can result in an unexpected resonant energy transfer between the QM and the internal molecular states when $J_0/(\hbar\omega)$ equals to 1 or 2, therefore should be avoided in a general control over polar molecules.

We find that high-fidelity molecular entanglement can still emerge in the presence of molecular QM. To show this, we introduce two high-fidelity gates. First, we find that the iSWAP gate can be generated by one microwave pulse of pulse area about $1.7\pi$, and the gate works with $J/(\hbar\Omega_\mu)>1/2$, which is in contrast to the standard molecular iSWAP gate with a $\pi$-wait-$\pi$ sequence where a high fidelity depends on a sufficiently small $J/(\hbar\Omega_\mu)$. Second, we introduce a two-qubit controlled phase gate with an arbitrarily desired phase realized by a pulse area $4\pi$. Both the one-pulse iSWAP gate and the two-qubit controlled phase gate can attain a high fidelity in the presence of QM-DDI coupling under typical experimental conditions.

\section*{acknowledgments}
We acknowledge the National Natural Science Foundation of China under Grants No. 12547103 and No. 12074300, and the Innovation Program for Quantum Science and Technology 2021ZD0302100 for support.

%


\end{document}